\documentclass[10pt,letterpaper]{article}

\usepackage[top=0.85in,left=2.75in,footskip=0.75in]{geometry}

\usepackage{amsmath,amssymb,mathtools}

\usepackage{dsfont}

\usepackage{changepage}

\usepackage{textcomp,marvosym}

\usepackage{cite}

\usepackage{nameref,hyperref}

\usepackage[right]{lineno}

\usepackage{booktabs}

\usepackage{makecell}

\usepackage{multirow}

\usepackage{caption}
\usepackage{subcaption}

\usepackage{bm}

\usepackage[nopatch=eqnum]{microtype}
\DisableLigatures[f]{encoding = *, family = * }

\usepackage[table]{xcolor}

\usepackage[english]{babel}
\usepackage[autostyle, english = american]{csquotes}
\MakeOuterQuote{"}

\usepackage{array}

\newcolumntype{+}{!{\vrule width 2pt}}

\newlength\savedwidth

\raggedright
\usepackage[aboveskip=1pt,labelfont=bf,labelsep=period,justification=raggedright,singlelinecheck=off]{caption}

\makeatletter
\renewcommand{\@biblabel}[1]{\quad#1.}
\makeatother

\usepackage{lastpage,fancyhdr,graphicx}
\usepackage{epstopdf}
\fancyheadoffset[L]{2.25in}
\fancyfootoffset[L]{2.25in}
\DeclarePairedDelimiter\autobracket{(}{)}
\newcommand{\br}[1]{\autobracket*{#1}}

\renewcommand{\Pr}{\text{Pr}}

\usepackage{dcolumn}
\newcolumntype{d}[1]{D{.}{.}{#1}}
\DeclareUnicodeCharacter{2212}{-}

\def\ie{{\em i.e.},\ }
\def\eg{{\em e.g.},\ }
\long\def\comment#1{}

\newcommand{\usecolor}{0}

\ifnum\usecolor=1
  
\else
  
\fi

\usepackage[inline]{trackchanges}
\addeditor{RG} 
\addeditor{CL}
\addeditor{SQ}  

\begin{document}
\vspace*{0.2in}

\begin{flushleft}
{\Large
\textbf\newline{A meaningful prediction of functional decline in amyotrophic lateral sclerosis based on multi-event survival analysis} 
}
\newline
\\
Christian Marius Lillelund*\textsuperscript{1}\textsuperscript{2},
Sanjay Kalra\textsuperscript{1}\textsuperscript{3}\textsuperscript{4},
Russell Greiner\textsuperscript{1}\textsuperscript{5},
The Pooled Resource Open-Access ALS Clinical Trials Consortium (PRO-ACT)\textsuperscript{\textdagger}
\\
\bigskip
\textbf{1} Department of Computing Science, University of Alberta, Edmonton, Alberta, Canada  
\\
\textbf{2} Department of Electrical and Computer Engineering, Aarhus University, Aarhus, Denmark  
\\
\textbf{3} Neuroscience and Mental Health Institute, University of Alberta, Edmonton, Alberta, Canada  
\\
\textbf{4} Division of Neurology, Department of Medicine, University of Alberta, Edmonton, Alberta, Canada  
\\
\textbf{5} Alberta Machine Intelligence Institute, University of Alberta, Edmonton, Alberta, Canada  
\bigskip

\textsuperscript{\textdagger}A complete list of members of the Pooled Resource Open-Access ALS Clinical Trials Consortium (PRO-ACT) can be found in the Acknowledgments.

*clillelund@ualberta.ca

\end{flushleft}
\section*{Abstract}
Amyotrophic lateral sclerosis (ALS) is a degenerative disorder of the motor neurons that causes progressive paralysis in patients. Current treatment options aim to prolong survival and improve quality of life. However, due to the heterogeneity of the disease, it is often difficult to determine the optimal time for potential therapies or medical interventions. In this study, we propose a novel method to predict the time until a patient with ALS experiences significant functional impairment (ALSFRS-R$\leq$2) for each of five common functions: speaking, swallowing, handwriting, walking, and breathing. We formulate this task as a multi-event survival problem and validate our approach in the PRO-ACT dataset ($N = 3\text{,}220$) by training five covariate-based survival models to estimate the probability of each event over the 500 days following the baseline visit. We then predict five event-specific individual survival distributions (ISDs) for a patient, each providing an interpretable estimate of when that event is likely to occur. The results show that covariate-based models are superior to the Kaplan-Meier estimator at predicting time-to-event outcomes in the PRO-ACT dataset. Additionally, our method enables practitioners to make individual counterfactual predictions -- where certain covariates can be changed -- to estimate their effect on the predicted outcome. In this regard, we find that Riluzole has little or no impact on predicted functional decline. However, for patients with bulbar-onset ALS, our model predicts significantly shorter time-to-event estimates for loss of speech and swallowing function compared to patients with limb-onset ALS (log-rank $p<0.001$, Bonferroni-adjusted $\alpha=0.01$). The proposed method can be applied to current clinical examination data to assess the risk of functional decline and thus allow more personalized treatment planning.

\section*{Introduction}

Amyotrophic lateral sclerosis (ALS) is a neurological disease that causes a gradual loss of upper and lower motor neurons in the central nervous system. Unfortunately, most patients will die within 2-5 years after onset~\cite{Morris2015}, primarily due to complications from respiratory failure~\cite{Talbot2009}. The underlying causes of the disease are not yet understood, and as there are no known cures, current treatments aim only to prolong survival and improve quality of life. Moreover, due to its heterogeneous and unpredictable nature, and the high variability in its progression rate and clinical phenotype~\cite{Feldman2022}, it is often difficult to determine the appropriate treatment. This also complicates decisions about the optimal timing of medical interventions, \eg non-invasive ventilation, or if a specific treatment can effectively slow the progression of the disease. There is therefore a significant need for tools that can predict disease progression to facilitate personalized treatment plans and improve patient care.

The ALS Functional Rating Scale-Revised (ALSFRS-R)~\cite{Cedarbaum1999} is the most widely used questionnaire to evaluate the progression of ALS. It contains 12 questions to assess functional capacity in patients, with a focus on bulbar, motor, and respiratory functions -- \eg swallowing, walking, and breathing, respectively. Each response is scored between 0 and 4, where 0 represents complete inability with regard to that function, and 4 represents normal function. As this questionnaire is administered to patients at multiple time points over the course of the disease, it provides a longitudinal view of functional decline across several areas. Based on this questionnaire, an interesting task is to predict the degree of functional decline a patient experiences during the course of the disease, rather than simply predicting when a patient is likely going to die~\cite{Kuan2023, VanDerBurgh2017}. Many previous projects have therefore attempted to predict the future ALSFRS-R score based on patient covariates~\cite{Ong2017, Vieira2022, Amaral2024, Jabbar2024, Turabieh2024}. 
Ong et al.~\cite{Ong2017} proposed a binary model to predict the functional decline class (\ie whether ALS progresses slowly or rapidly in a specific patient) after baseline (trial entry), based on several patient characteristics. Their model was able to predict slow or rapid decline, but did not predict when functional decline would occur over time.
Similarly, Jabbar et al.~\cite{Jabbar2024} proposed a method to predict disease progression as fast or slow, but as in \cite{Ong2017}, their model could only give a yes/no answer to whether a patient would experience functional decline over time. Amaral et al.~\cite{Amaral2024} applied several clustering algorithms to clinical patient records, which revealed four separate groups regarding disease progression. Their method could assign newly-diagnosed patients to a progression group, but did not predict when a patient would experience functional decline based on their covariates. 
Finally, Vieira et al.~\cite{Vieira2022} trained two machine learning models to predict a patient's future ALSFRS-R score: a voice model based on speech recordings and a movement model based on accelerometer measurements. These models showed strong predictive performance, but similar to Amaral et al.~\cite{Amaral2024} and others, they predicted future ALSFRS-R scores rather than estimating the time until functional decline would occur. Moreover, disease progression in ALS is neither constant nor linear, as those studies implicitly assumed, but can vary substantially over time and between individuals~\cite{Kjaeldgaard2021}.

Based on these shortcomings, we propose a novel method that can predict when an ALS patient will experience each of five common types of functional decline, based on the ALSFRS-R protocol (see Table \ref{tab:alsfrs_definitions}): significant difficulty ($\leq$ 2) in each of \emph{Speaking}, \emph{Swallowing}, \emph{Handwriting}, \emph{Walking}, and \emph{Dyspnea}. These five events are distinct but not mutually exclusive and likely share some information in their covariates; it may be that a patient (\eg Mr. Smith), with limb-onset ALS, will experience difficulty walking and writing earlier than another patient (\eg Mr. Johnson), who has bulbar-onset ALS. On the other hand, Mr. Johnson may have difficulty speaking and swallowing solid foods before Mr. Smith. As a newly-diagnosed patient with ALS, our approach will help answer important questions, such as:

\begin{itemize}
\item How much longer will I be able to speak clearly enough to be understood? 
\item How much longer will I be able to eat most solid foods without difficulty?
\item How much longer will I be able to write a shopping list that my wife can read?
\item How much longer will I be able to walk at a normal pace and keep my balance?
\item How much longer will I be able to breathe normally during daily activities?
\end{itemize}

In practice, we formulate the problem as a multi-event survival problem: given an instance (\eg a description of Mr. Smith at a "baseline" point in time), our method predicts the time until said Mr. Smith will experience each of the events listed above. We present each prediction as an individual survival distribution (ISD)~\cite{Haider2020}, which provide the probability of each event at each future time point, after the baseline visit. We describe each patient by the values of a set of covariates such as age, site of onset, forced vital capacity (FVC), and others, which are recorded in clinical records obtained at or before the baseline visit. To our knowledge, no existing method can explicitly predict the timing of distinct individual events in ALS. Fig.~\ref{fig:als_model} shows an outline of the proposed method. The source code is publicly available at: \url{https://github.com/thecml/FunctionalALS}.

\begin{figure}[!htbp]
\vspace{0.5cm}
\begin{adjustwidth}{-2.25in}{0in}
\centering
\textbf{[Fig. 1]}
\caption{Outline of the proposed method. Historical data about ALS patients form a patient dataset $\mathcal{D}$ with $N$ training instances and $d$ covariates, and the time to event (filled dot) or censoring (hollow dot) from a patient's first visit -- the "baseline time". Censoring indicates that the event of interest was not observed for a patient within the study period, so their exact event time remains unknown. We consider five separate but related events, \ie \emph{Speech}, \emph{Swallowing}, \emph{Handwriting}, \emph{Walking}, and \emph{Dyspnea}. We use the recorded covariates (taken at the baseline time) and event information to train a survival model $\mathcal{M}$ that can accurately estimate the individual survival distribution (ISD) of each of these five events, for a novel patient $\bm{x}_{i}$, denoted as $\hat{S}^{(i)}$. These ISDs give the probability of each of these five events occurring after $t$ days after the baseline visit, for all $t>0$. They can also be used to estimate the time to event for this $\bm{x}_{i}$ patient, for example, when the survival curve intersects the dashed horizontal line at 50\%, which is called the median survival time, for each of the five events.}
\label{fig:als_model}
\end{adjustwidth}
\end{figure}

\begin{table}[!htbp]
\begin{adjustwidth}{-2.25in}{0in}
\centering
\caption{The definition of the five included functional assessments from the ALS Functional Rating Scale-Revised (ALSFRS-R)~\cite{Cedarbaum1999} and their scores. Each score ranges from 0 to 4, where 0 represents complete inability with regard to the function, and 4 represents normal function. For each patient, our method predicts the time until that patient will experience significant difficulty (ALSFRS-R $\leq$ 2) for that specific assessment.}
\label{tab:alsfrs_definitions}
\begin{tabular}{lll c}
\toprule
\makecell[l]{Domain} & \makecell[l]{Assessment} & \makecell[l]{Description} & \makecell{Score} \\
\midrule
\makecell[l]{Bulbar} & \makecell[l]{Speech} & \makecell[l]{Ability to speak clearly and be understood.} & \makecell{4 = Speaks clearly;\\2 = Intelligible with repeating;\\0 = Loss of useful speech} \\
\midrule
\makecell[l]{Bulbar} & \makecell[l]{Swallowing} & \makecell[l]{Ability to swallow without choking.} & \makecell{4 = Normal eating habits;\\2 = Difficulty with some foods;\\0 = Needs feeding tube} \\
\midrule
\makecell[l]{Fine motor} & \makecell[l]{Handwriting} & \makecell[l]{Ability to write or type.} & \makecell{4 = Normal writing;\\2 = Not all words legible;\\0 = Unable to grip pen} \\
\midrule
\makecell[l]{Gross motor} & \makecell[l]{Walking} & \makecell[l]{Ability to walk without assistance.} & \makecell{4 = Walks normally;\\2 = Walks with assistance;\\0 = Cannot walk} \\
\midrule
\makecell[l]{Respiratory} & \makecell[l]{Dyspnea} & \makecell[l]{Shortness of breath.} & \makecell{4 = None;\\2 = Moderate shortness of breath;\\0 = Severe shortness of breath} \\
\bottomrule
\end{tabular}
\end{adjustwidth}
\end{table}

\section*{Materials and methods}

\subsection*{Clinical data and event annotation}

For our empirical analyses, we use the publicly-available Pooled Resources Open-Access Clinical Trials (PRO-ACT)\footnote{Data used in the preparation of this article were obtained from the Pooled Resource Open-Access ALS Clinical Trials (PRO-ACT) Database. The data available in the PRO-ACT Database have been volunteered by PRO-ACT Consortium members. The dataset was obtained on October 1st, 2024.}~\cite{atassi_proact_2014} dataset, which is the largest ALS dataset in the world. It includes patient demographics, lab and medical records, as well as family histories, of over 11,600 ALS patients from 23 clinical trials (see Table \ref{tab:patient_demographics}). In each clinical trial, patients follow a schedule of visits until the end of the study or until they become too ill to participate. Covariates recorded at each visit include the total ALSFRS-R score, age, sex, region of onset (\eg limb, bulbar), the mean FVC score, time since diagnosis, disease progression rate\footnote{Defined as (48 - the total ALSFRS-R score)/(days since diagnosis/30).}, and whether or not the patient is taking Riluzole. In this study, we only use covariates recorded at the baseline visit for each individual patient.

\begin{figure*}[!htbp]
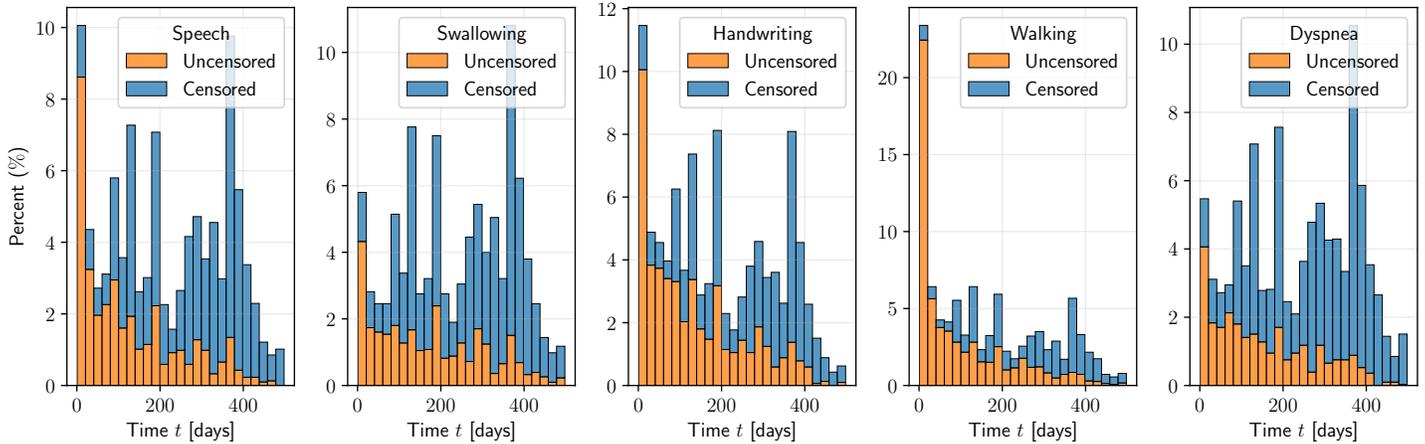

\begin{adjustwidth}{-2.25in}{0in}
\centering
\textbf{[Fig. 2]}
\caption{Distribution of uncensored and censored times in the PRO-ACT dataset for the five events.}
\label{fig:event_distribution}
\end{adjustwidth}
\end{figure*}

To define our events of interest, each patient's disease state is assessed by a clinician at every visit using the ALSFRS-R scale. We focus on five specific events: \emph{Speech}, \emph{Swallowing}, \emph{Handwriting}, \emph{Walking}, and \emph{Dyspnea}. To annotate our dataset, we define an event as having occurred the first time a patient scores 2 or fewer in that specific assessment during any follow-up visit following their baseline visit. These five events can occur in any order and are not mutually exclusive. There are several types of censoring, including death, being unable to perform the assessment or leaving the study prematurely. Censoring thus represents incomplete event information, where the true event time is only known to exceed the observed follow-up time. We use a maximum follow-up time of 500 days to focus on the early stages of the disease and because data become sparse with high censoring afterward. We consider patients who score 2 or fewer on any of the five assessments at baseline as having already had the event, indicating that the loss of function occurred before the start of the trial, and exclude these patients from the study. We also exclude patients with no recorded ALSFRS-R history. After applying these criteria and accounting for censoring, the final dataset comprises 3,220 patients with observed or right-censored event times. The code for loading and annotating the data is available in the source code repository. Table~\ref{tab:dataset_overview} provides an overview of the dataset. Fig.~\ref{fig:event_distribution} shows the distribution of event and censoring times. A list of covariates is available in the Supplement (\nameref{S1_Table}).

\begin{table}[!ht]
\centering
\caption{Patient demographics in the PRO-ACT \cite{atassi_proact_2014} dataset.}
\label{tab:patient_demographics}
\begin{tabular}{ll}
\toprule
\textbf{Demographic/Data} & \makecell{\textbf{PRO-ACT (N=3,220)} \\ Pct. or mean ($\pm$ SD)} \\
\midrule
Age (years) & 55.8 (11.6) \\
Height (cm) & 172 (9.5) \\
Weight (kg) & 76.3 (14.6) \\
Body mass index (BMI) & 25.7 (4) \\
\% Female & 32.2 \\
\% Caucasian & 94.6 \\
Site of Onset; \% Limb, \% Bulbar & 43.3, 13.4 \\
Baseline ALSFRS-R Score & 39.4 (5) \\
Time in study (days) & 271.4 (122.7) \\
\% Riluzole & 75.7 \\
\bottomrule
\end{tabular}
\end{table}

\begin{table}[!ht]
\centering
\caption{Overview of the dataset. "SP", "SW", "HA", "WA" and "DY", are the \emph{Speech}, \emph{Swallowing}, \emph{Handwriting}, \emph{Walking} and \emph{Dyspnea} events, respectively. Event distribution indicates the percentage of uncensored instances of each type in the dataset. $N$: sample size, $d$: number of covariates, $K$: number of event types, $T$: observed time (of the uncensored instances) for any event in days.}
\begin{tabular}{ccccccccc}
\toprule
Dataset & N & d & \makecell{K} & \makecell{T$_{min}$} & \makecell{T$_{max}$} & \makecell{T$_{mean}$} & \makecell{Event distribution (\%):} \\
\midrule
PRO-ACT & 3,220 & 8 & 5 & 1 & 498 & 130 & \makecell{SP: 37.8, SW: 31.9\\
HA: 50.3, WA: 60.6\\ DY: 27.0} \\
\bottomrule 
\end{tabular}
\label{tab:dataset_overview}
\end{table}

\subsection*{Survival analysis and notation}

Survival analysis models the time until some event of interest occurs. This event can either be observed or censored -- the latter typically happens if the patient is still alive at the end of the study or decides to drop out~\cite[Ch. 11]{gareth_introduction_2021}. Survival analysis has become an important tool for understanding diseases and predicting outcomes, which helps medical researchers evaluate the significance of prognostic factors in applications like comatose after cardiac arrest~\cite{Shen2023}, liver transplantation~\cite{Andres2018} or fall risk~\cite{lillelund_prognosis_2024}. In ALS, survival analysis has been used mainly to predict risk scores with respect to death~\cite{Kjaeldgaard2021, Kuan2023, Lajoie2024}.

We briefly introduce the notation: Let $\mathcal{D} = \left\lbrace \br{\bm{x}^{(i)}, \bm{t}^{(i)}, \bm{\delta}^{(i)}} \right\rbrace_{i=1}^N$ be the dataset, where for each $i$, $\bm{x}^{(i)} \in \mathbb{R}^{d}$ is a $d$-dimensional vector of covariates. We consider discrete times, where $\bm{t}^{(i)} \in \{1,2,\ldots,T_{\text{max}}\}^{K}$ is a vector of observed times for the $i$-th patient for $K$ different events, each at one of these $T_{\text{max}}$ possible times, and $\bm{\delta}^{(i)} \in \{0,1\}^{K}$ is a vector of event indicators for each event. Moreover, let $e_{k}^{(i)} \in \{1,2,\ldots,T_{\text{max}}\}$ denote the event time and let $c_{k}^{(i)} \in \{1,2,\ldots,T_{\text{max}}\}$ denote the censoring time for the $i$-th instance for event $k$ over the event horizon, thus we have $t_{k}^{(i)} = e_{k}^{(i)}$ if $\delta_{k}^{(i)} = 1$, otherwise, $t_{k}^{(i)} = c_{k}^{(i)}$ if $\delta_{k}^{(i)} = 0$. A survival model predicts the probability that some event $k$ occurs at time $T$ later than $t$, \ie the so-called survival or event probability, which is denoted $S\br{t} = \Pr\br{T>t} = 1-\Pr\br{t\leq T}$. In multi-event survival analysis, we want to predict the probability of different events occurring over time. For each patient $i$, this results in a matrix of survival functions, $\bm{S}^{(i)}=\{\bm{s}_{1}^{(i)}, \bm{s}_{2}^{(i)}, \ldots, \bm{s}_{K}^{(i)}\}$. Here, $\bm{s}_{k}^{(i)}$ represents the probabilities that patient $i$ does not experience event $k$ up to each time point $t$, \ie $\bm{s}_{k}^{(i)} = (P(e_{k}^{(i)} > 1), P(e_{k}^{(i)} > 2), \ldots, P(e_{k}^{(i)} > T_{\text{max}}))$. See Fig. \ref{fig:als_model} for a visual overview.

\section*{Experiments and Results}

\subsection*{Setup and preprocessing}

We follow Sechidis et al.~\cite{Sechidis2011} and first split the dataset into a training, validation and test set by 70\%, 10\%, and 20\% using a stratified procedure. This ensures that the event times and censoring rates are consistent across the three sets and the $K$ events. The stratification procedure uses a random seed (0-9). After splitting the data, we impute missing values using the sample mean for real-valued covariates or the mode for categorical covariates based on the training set only. A list of missing values is available in the Supplement (\nameref{S2_Table}). After imputing missing values, we encode categorical covariates using a one-hot encoding strategy and apply a $z$-score data normalization to speed up the training process and improve stability. No outlier detection was performed.

Our evaluation procedure consists of training five survival models on the training set and evaluating them on the holdout test set. For evaluation fairness, we do not perform hyperparameter optimization for specific models, but instead configure all models with sensible default parameters. A list of hyperparameters is available in the Supplement (\nameref{S3_Table}). For deep learning models, we use a single hidden layer with 32 nodes. For methods that support it, we use early stopping during training to terminate the process if the validation loss does not improve for 10 consecutive epochs. All data preprocessing and experiments were implemented in Python 3.9. A list of software packages used is available in the source code repository.

\subsection*{Survival models}

To evaluate both single-event and multi-event survival modeling approaches, we distinguish between models that predict the time to a single event and those that jointly model multiple event types. As single-event models, we implement the traditional Cox Proportional Hazards (CoxPH) model using a linear estimator~\cite{cox_regression_1972}, Random Survival Forests (RSF)~\cite{ishwaran_random_2008}, DeepSurv~\cite{katzman_deepsurv_2018} and MTLR~\cite{yu_learning_2011}. As a multi-event model, we implement the Multi-Event Network for Survival Analysis (MENSA) model~\cite{lillelund_mensa_2025}. To explain these models: \textbf{CoxPH} is a popular semiparametric method to fit a regression model to survival data. By adopting a multiplicative form for the contribution of several covariates to each individual's event time, the CoxPH model is a powerful yet simple tool for assessing the simultaneous effect that different covariates have on the event times. CoxPH assumes proportional hazards, which means that the effect of covariates on the hazard function is constant over time. \textbf{RSF} extends decision trees for survival analysis and recursively splits training data based on some survival-specific criterion, \eg log-rank splitting, with the goal of maximizing separation between event times between nodes. RSF can model nonlinear relationships between the covariates and the event, and does not assume proportional hazards. \textbf{DeepSurv} is a multilayer perceptron (MLP) based on the CoxPH model, where the risk score is a nonlinear function of the covariates. DeepSurv assumes proportional hazards. \textbf{MTLR} is a discrete survival model that estimates the individual survival function as a multinomial logistic regression model and does not assume proportional hazards. Lastly, \textbf{MENSA} is an MLP that learns the individual survival distribution for $K$ events as a mixture of Weibull distributions with a shared covariate layer to model dependencies between events. MENSA does not assume proportional hazards.

\subsection*{Evaluation metrics}

To comprehensively assess the survival models, we select a range of evaluation metrics that capture three key aspects of model performance: discrimination, absolute/squared error, and calibration. Discrimination, usually measured by the concordance index (CI), is relevant when we want to identify which patients are likely to experience functional decline earlier versus later at the group level. Squared or absolute error metrics, like the mean absolute error (MAE), are used to assess how accurately the model predicts the time of functional impairment for individual patients. Finally, calibration measures how well the predicted probabilities match the observed frequencies of events at specific time points. For example, if a model predicts that a patient has an 80\% chance of not experiencing significant functional decline before 100 days, then among 100 similar patients, about 80 should not experience significant functional decline by that time. We use the following metrics to assess discrimination, absolute/squared error, and calibration:

\textbf{CI:} Harrell's CI~\cite{Harrell1982} measures discrimination performance by calculating the proportion of concordant pairs among all comparable pairs given predicted risk scores. A pair is considered comparable if the event order can be determined. Specifically, for a comparable pair $\{i, j\}$ with event times $e_i < e_j$ and $\delta_i = 1$ (indicating that $i$ experienced the event), we calculate risk scores $\hat{\eta}_i$ and $\hat{\eta}_j$ as the negative median survival times.
The median survival time is the time point at which the estimated survival probability drops to 0.5. If the predicted risk score for $j$ is greater than that for $i$ at the time $i$ experienced the event (while $j$ remains event-free), the pair is concordant. A CI of 0.5 indicates chance-level ranking. We also report Uno's CI~\cite{Uno2011}, which is a CI that uses a technique to correct for the bias introduced by censoring.

\textbf{BS/IBS:} The Brier Score (BS) calculates the squared error between the predicted probability of the event and the Heaviside step function of the observed event. The integrated Brier score~\cite{graf1999assessment} (IBS) aggregates Brier scores over multiple time points to provide a single measure of the squared error.

\textbf{MAE:} The MAE calculates the average absolute difference between the true event time and the predicted event time. Given an ISD, $S\br{t\mid\bm{x}_{i}} = \text{Pr}\br{T>t\mid\bm{x}_{i}}$, we calculate the predicted event time $\hat{t}_i$ as the median survival time~\cite{qi_survivaleval_2024}. Since our dataset contains censoring, we calculate the margin MAE (mMAE) as proposed by~\cite{Haider2020}, which assigns a "best-guess" estimate to each censored patients using the Kaplan-Meier (KM)~\cite{kaplan_nonparametric_1958} estimator.

\textbf{D-calibration:} Distribution calibration~\cite{Haider2020}, also known as D-Cal, measures how well the predicted survival function, $\hat{S}(t)$, is calibrated for each event. D-calibration assesses this using a Pearson's chi-squared test with $\alpha = 0.05$. For any probability interval $[a, b] \in [0, 1]$, we define $D_m(a, b)$ as the group of patients in the dataset $D$ whose predicted probability of an event is in the interval $[a, b]$~\cite{qi_effective_2023}. A model is D-calibrated if the proportion of patients $|D_m(a, b)|/|D|$ is statistically similar to the amount $b-a$. Intuitively, if a model is well-calibrated, the predicted probabilities match observed event frequencies -- for instance, about 20\% of patients predicted to have a 0.2 event probability should actually experience the event. In our experiments, we report the number of times each model is D-calibrated (at $p>0.05$) for each event across 10 experiments.

\subsection*{Model results}

\begin{figure}[!ht]
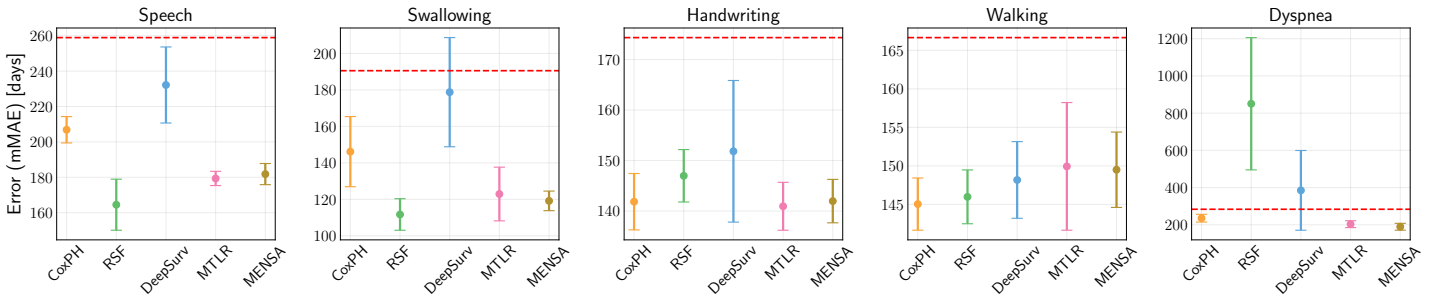

\begin{adjustwidth}{-2.25in}{0in}
\centering
\textbf{[Fig. 3]}
\caption{The mMAE (in days) as a function of covariate-based models (x-axis) in the PRO-ACT test set, with error bars representing empirical 95\% confidence intervals. The horizontal dashed line is the KM estimator. Lower is better.}
\label{fig:km_mae_vs_model}
\end{adjustwidth}
\end{figure}

We present our empirical results (Table \ref{tab:model_results} and Fig. \ref{fig:km_mae_vs_model}) in terms of the aforementioned aspects with respect to model performance: discrimination, absolute/squared error, and calibration. Below, we discuss the results and reflect on their implications.

\textbf{Discrimination:} Good discrimination performance is important if we need to rank patients by their disease severity to prioritize access to limited clinical resources, for example, portable ventilators. Across all models, we see the highest Harrell's and Uno's C-indices for the \emph{Speech} (approx. 74–75\% and 70–71\%, respectively) and \emph{Swallowing} (approx. 75–76\% and 70–71\%, respectively) events. On the other hand, functional decline with respect to limb functions is harder to discriminate, with lower C-indices (approx. 67–69\%) for the \emph{Handwriting} and \emph{Walking} events. Although the \emph{Dyspnea} event typically has moderate CI values (approx. 66–68\%), RSF achieves a Harrell's CI of 78.23\% and a Uno's CI of 78.95\% for this event, making it the best model in this case. DeepSurv and MTLR also demonstrate consistently good discrimination performance across events, and MENSA matches or slightly exceeds these for the \emph{Speech} and \emph{Swallowing} events, though its performance dips slightly for the \emph{Walking} and \emph{Dyspnea} events.

\textbf{Absolute/squared error:} Accurate time-to-event estimates or survival probabilities are important when a decision or intervention depends on the precise timing of the event. Across all models, the \emph{Swallowing} event consistently exhibits some of the lowest IBS (approx. 12-13) and mMAE (approx. 110-150) numbers, meaning that this event is the easiest to predict the timing of. For the \emph{Dyspnea} event, the IBS is also moderate (11.9-13.8), but the RSF model is notable with a low IBS but a relatively high mMAE (approx. 850$\pm$471 days). In contrast, other models produce much more moderate mMAE numbers for this event (approx. 200-400 days), as seen with the MENSA model. This highlights an important point: simply calculating the squared error of the predicted survival probabilities does not fully capture a model's ability to predict the timing of an event. MENSA achieves reasonable mMAE numbers (approx. 120-190 days) in all cases, except for the \emph{Speech} event (181$\pm$8 days), however, and tends to perform particularly well for the \emph{Swallowing} event, similar to MTLR.

Fig. \ref{fig:km_mae_vs_model} shows the prediction error in days between the population-level KM estimator~\cite{kaplan_nonparametric_1958} and the covariate-based models. The KM estimate is calculated per event on observed and censored event times, without considering individual differences (\ie covariates). In all five events, the covariate-based models generally give superior time-to-event estimates compared to the KM method, with some variation between models. This suggests that incorporating patient-specific information can lead to more accurate time-to-event predictions when predicting functional decline, rather than simply relying on a population-level estimate.

\textbf{Calibration:} Good calibration performance is important when we have to use the actual predicted probabilities to guide our decision-making. For example, underestimating the risk of respiratory decline or bulbar involvement may result in mistimed interventions or suboptimal treatment choices. Across all models, the \emph{Walking} event consistently poses a calibration challenge, as predicted event probabilities generally deviate from the observed event frequencies here (\ie the models are sometimes not calibrated). However, RSF achieves perfect D-calibration (10/10) across all events and experiments, demonstrating robust and consistent calibration performance. DeepSurv and MTLR also perform well, with most experiments achieving D-calibration. MENSA achieves perfect calibration for the \emph{Swallowing} and \emph{Dyspnea} (10/10) events, but shows slightly poorer calibration for the \emph{Speech} (9/10) and \emph{Handwriting} (9/10) events, and especially the \emph{Walking} event (only 6/10).

\begin{table}[!ht]
\begin{adjustwidth}{-2.25in}{0in}
\centering
\caption{Mean ($\pm$ SD.) prediction performance, averaged over 10 independent experiments. D-calibration counts the number of times the respective model was D-calibrated. Harrell's CI, Uno's CI and IBS results are multiplied by 100.}
\label{tab:model_results}
\begin{tabular}{@{}ccllllll@{}}
\toprule
\multirow{2}{*}{\makecell{Model}} &
\multirow{2}{*}{\makecell{Event}} &
\multirow{2}{*}{\makecell{Harrell's CI $\uparrow$}} &
\multirow{2}{*}{\makecell{Uno's CI $\uparrow$}} &
\multirow{2}{*}{\makecell{IBS $\downarrow$}} &
\multirow{2}{*}{\makecell{mMAE $\downarrow$}} &
\multirow{2}{*}{\makecell{D-Cal $\uparrow$}} \\ 
 & & & & & \\
\midrule
\multirow{5}{*}{\makecell{CoxPH \\ \cite{cox_regression_1972}}}
& Speech & 74.52$\pm$1.39 & 70.44$\pm$3.34 & 14.41$\pm$0.43 & 206.85$\pm$9.88 & (10/10) \\
& Swallowing & 75.75$\pm$1.53 & 70.96$\pm$4.58 & 12.58$\pm$0.72 & 146.15$\pm$25.50 & (10/10) \\
& Handwriting & 67.57$\pm$0.92 & 63.75$\pm$2.29 & 16.73$\pm$0.70 & 141.85$\pm$7.40 & (10/10) \\
& Walking & 68.67$\pm$1.20 & 67.15$\pm$1.40 & 16.60$\pm$0.50 & 145.04$\pm$4.50 & (7/10) \\
& Dyspnea & 68.54$\pm$1.71 & 66.99$\pm$6.30 & 13.79$\pm$0.41 & 235.46$\pm$27.57 & (10/10) \\
\cmidrule(lr){1-6}
\multirow{5}{*}{\makecell{RSF \\ \cite{ishwaran_random_2008}}}
& Speech & 75.05$\pm$1.13 & 70.08$\pm$2.75 & 14.69$\pm$0.44 & 164.51$\pm$19.12 & (10/10) \\
& Swallowing & 75.53$\pm$1.46 & 69.65$\pm$2.98 & 13.02$\pm$0.74 & 111.71$\pm$11.46 & (10/10) \\
& Handwriting & 67.38$\pm$1.39 & 63.70$\pm$2.05 & 16.87$\pm$0.57 & 146.97$\pm$6.88 & (10/10) \\
& Walking & 68.44$\pm$1.47 & 66.68$\pm$1.54 & 16.78$\pm$0.56 & 145.98$\pm$4.63 & (10/10) \\
& Dyspnea & 78.23$\pm$1.07 & 78.95$\pm$4.01 & 11.89$\pm$0.33 & 850.45$\pm$471.39 & (10/10) \\
\cmidrule(lr){1-6}
\multirow{5}{*}{\makecell{DeepSurv \\ \cite{katzman_deepsurv_2018}}}
& Speech & 74.18$\pm$3.05 & 70.77$\pm$3.57 & 14.36$\pm$0.89 & 232.15$\pm$28.48 & (10/10) \\
& Swallowing & 75.61$\pm$1.37 & 70.67$\pm$5.52 & 12.48$\pm$0.65 & 178.77$\pm$39.68 & (10/10) \\
& Handwriting & 67.12$\pm$1.08 & 63.72$\pm$2.15 & 16.81$\pm$0.84 & 151.83$\pm$18.59 & (9/10) \\
& Walking & 69.53$\pm$0.93 & 68.44$\pm$1.11 & 16.23$\pm$0.51 & 148.17$\pm$6.60 & (8/10) \\
& Dyspnea & 71.72$\pm$3.99 & 70.61$\pm$7.60 & 13.11$\pm$0.61 & 384.96$\pm$284.17 & (10/10) \\
\cmidrule(lr){1-6}
\multirow{5}{*}{\makecell{MTLR \\ \cite{yu_learning_2011}}}
& Speech & 74.22$\pm$1.67 & 70.17$\pm$3.23 & 13.95$\pm$0.44 & 179.35$\pm$5.34 & (10/10) \\
& Swallowing & 75.13$\pm$1.43 & 70.63$\pm$5.41 & 12.34$\pm$0.59 & 122.93$\pm$19.52 & (10/10) \\
& Handwriting & 67.32$\pm$1.11 & 63.71$\pm$2.42 & 16.72$\pm$0.66 & 140.94$\pm$6.29 & (9/10) \\
& Walking & 68.97$\pm$1.16 & 67.71$\pm$1.06 & 16.30$\pm$0.51 & 149.93$\pm$10.99 & (9/10) \\
& Dyspnea & 66.54$\pm$2.22 & 65.90$\pm$6.85 & 13.50$\pm$0.50 & 203.36$\pm$24.57 & (10/10) \\
\cmidrule(lr){1-6}
\multirow{5}{*}{\makecell{MENSA \\ \cite{lillelund_mensa_2025}}}
& Speech & 74.21$\pm$1.11 & 70.56$\pm$3.47 & 14.53$\pm$0.41 & 181.81$\pm$7.92 & (9/10) \\
& Swallowing & 74.94$\pm$1.36 & 70.85$\pm$5.20 & 12.61$\pm$0.56 & 119.16$\pm$7.12 & (10/10) \\
& Handwriting & 66.46$\pm$0.90 & 63.80$\pm$2.65 & 17.10$\pm$0.66 & 141.97$\pm$5.70 & (9/10) \\
& Walking & 68.63$\pm$1.29 & 66.94$\pm$1.50 & 17.04$\pm$0.61 & 149.50$\pm$6.49 & (6/10) \\
& Dyspnea & 66.56$\pm$1.77 & 64.40$\pm$7.07 & 13.78$\pm$0.30 & 189.43$\pm$24.28 & (10/10) \\
\bottomrule
\end{tabular}
\end{adjustwidth}
\end{table}

\subsection*{Individual disease predictions}

\begin{figure}[!htbp]
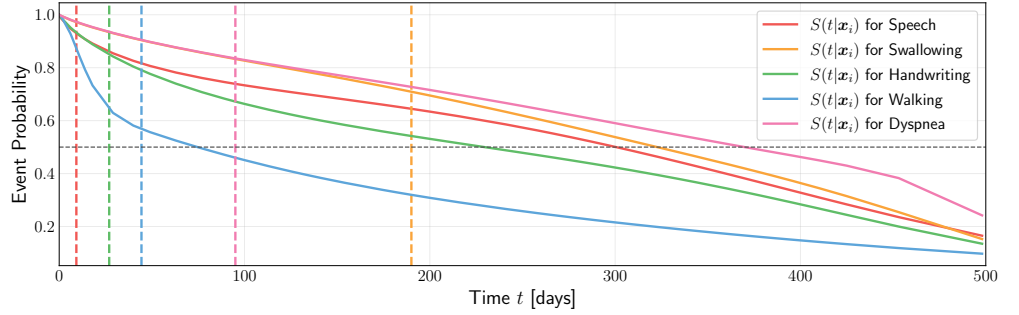
   
\centering
\textbf{[Fig. 4]}
\caption{Predicted ISDs for Mr. Smith ($i$). The point where the survival curve intersects the dashed horizontal line at 50\% indicates the predicted time of event -- so this predicts, for example, \emph{Walking} at approximately 80 days, \emph{Handwriting} at 240 days, and \emph{Speech} at 300 days. The dashed vertical lines are Mr. Smith's actual time of event for the respective events -- so, for example, \emph{Speech} at approximately 7 days, \emph{Handwriting} at 35 days, and \emph{Walking} at 45 days.}
\label{fig:proact_isd}
\end{figure}

Our method can predict a patient's disease trajectory as $K$ individual survival distributions, where $K$ is the number of events (here, $K=5$). Fig. \ref{fig:proact_isd} shows such ISDs, from baseline ($t=0$) to 500 days later, for the $i$-th patient. As an example, let us return to our patient, Mr. Smith. He is a 72-year-old male, who had been diagnosed with limb-onset ALS prior to study entry and had experienced difficulty speaking and weakness in his legs, but all five of his scores were above 2, so that he could participate in the study. Mr. Smith's covariates are available in the Supplement (\nameref{S4_Table}). The neurologist predicted that talking and walking difficulties would be among Mr. Smith’s earliest functional impairments, and our method confirmed her suspicion. From the outset, Mr. Smith was provided with a walking aid and a personalized long-term care plan that included therapy, support systems, and home modifications. Shortly after his hospital visit, he had to repeat many of his words to be understood and relied on assistance from a cane or his wife to walk. Meanwhile, handwriting also became challenging for him, but he did not experience shortness of breath or difficulty swallowing whole foods until much later.

\subsection*{Individual counterfactual predictions}

Our method can also make individual counterfactual predictions~\cite{Prosperi2020}, that is, alternative scenarios in which certain patient covariates are changed to see their (predicted) effect on the probability of the event. Such predictions can give novel insights into disease mechanisms and assess treatment options on an individual level. In the following figures, we have used MENSA~\cite{lillelund_mensa_2025} to make individual counterfactual predictions based on our PRO-ACT dataset. These provide insights into the relationship between common disease predictors and functional decline, and ask how changing certain covariates affects disease outcomes. Again, we use our patient, Mr. Smith, as the instance $i$. Additionally, we separate all patients in the test set into groups by the same covariate we used for the individual patient and employ the log-rank test~\cite{Mantel1966} to measure the statistical difference between the groups' event times. We use a Bonferroni correction to correct for multiple comparisons by dividing the significance level ($\alpha = 0.05$) by the number of comparisons ($n=5$), \ie the significance level is then $\alpha_{adjusted} = 0.05/5 = 0.01$. For the remainder of this section, we will discuss our counterfactual predictions.

\begin{figure}[!ht]
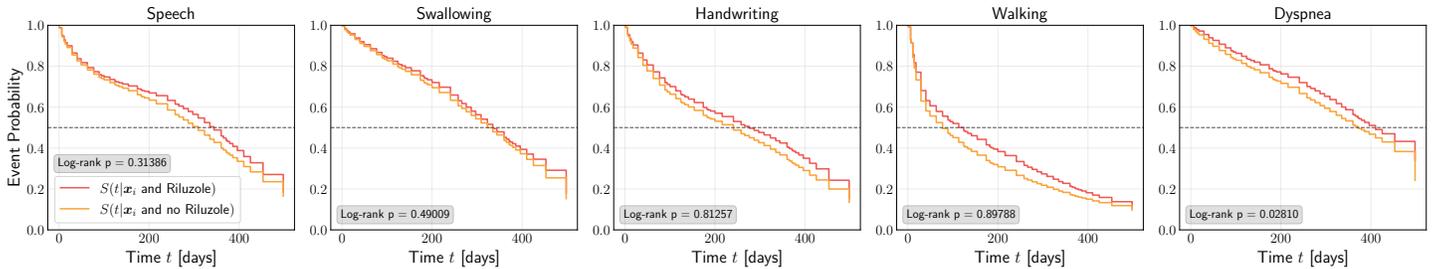

\begin{adjustwidth}{-2.25in}{0in}
\centering
\textbf{[Fig. 5]}
\caption{Predicted ISDs for Mr. Smith based on his use of Riluzole.}
\label{fig:proact_isd_riluzole}
\end{adjustwidth}
\end{figure}

As a starting point, many newly diagnosed patients with ALS need to decide whether to take certain pharmaceutical drugs. Currently, there are three approved drugs by the US Food and Drug Administration (FDA) to treat ALS, the most common being Riluzole. Riluzole works by blocking the release of glutamate, which delays the onset of ventilator dependence or tracheostomy in some people and may increase overall survival by two to three months~\cite{Jaiswal2019}. However, the drug can cause significant side effects, including nausea, anorexia, and diarrhea~\cite{Kjaeldgaard2021}. Therefore, it is important to determine whether the drug will be effective -- \ie significantly delay functional decline in this regard -- for each individual, in order to decide whether that person should receive the treatment. Upon diagnosis, Mr. Smith decided not to take Riluzole, but what if he instead had decided to take it. Fig. \ref{fig:proact_isd_riluzole} shows the predicted ISDs for each of the five events using Mr. Smith's covariates ($\bm{x}_i$), with the Riluzole covariate being yes (resp., no) at the baseline visit. The two survival curves have nearly the same slopes, with Riluzole only being marginally above non-Riluzole, and nearly identical estimated event times. There are no statistical differences in event times between the groups either ($p>0.01$). As an example, for the \emph{Speech} event, this model predicts that taking Riluzole would delay the onset of speaking difficulties by approximately 20 days.

\begin{figure}[!ht]
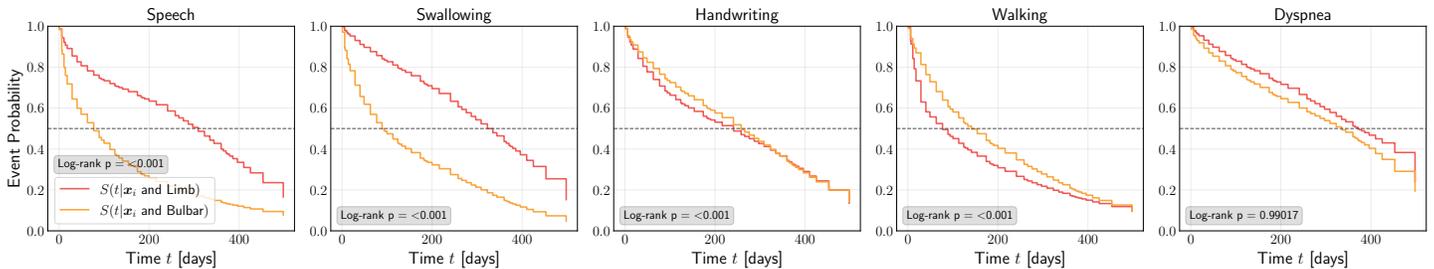

\begin{adjustwidth}{-2.25in}{0in}
\centering
\textbf{[Fig. 6]}
\caption{Predicted ISDs for Mr. Smith based on whether he has limb-onset or bulbar-onset.}
\label{fig:proact_isd_soo}
\end{adjustwidth}
\end{figure}

In ALS, the site of onset refers to the specific part of the body where the initial symptoms began. ALS can affect different regions, and the site of onset can influence the progression of the disease. The two most common sites of onset are bulbar-onset and limb-onset, the former being associated with faster disease progression in general~\cite{Jabbar2024}. Mr. Smith was diagnosed with limb-onset ALS, but what if instead he had been diagnosed with bulbar-onset ALS. Fig. \ref{fig:proact_isd_soo} shows the predicted ISDs for Mr. Smith based on whether he has limb-onset or bulbar-onset. We see significant differences in the predictions for the \emph{Speech} and \emph{Swallowing} events, indicating a worse prognosis for Mr. Smith if he has bulbar-onset, but more agreement between the curves for the \emph{Handwriting}, \emph{Walking} and \emph{Dyspnea} events.

\begin{figure}[!ht]
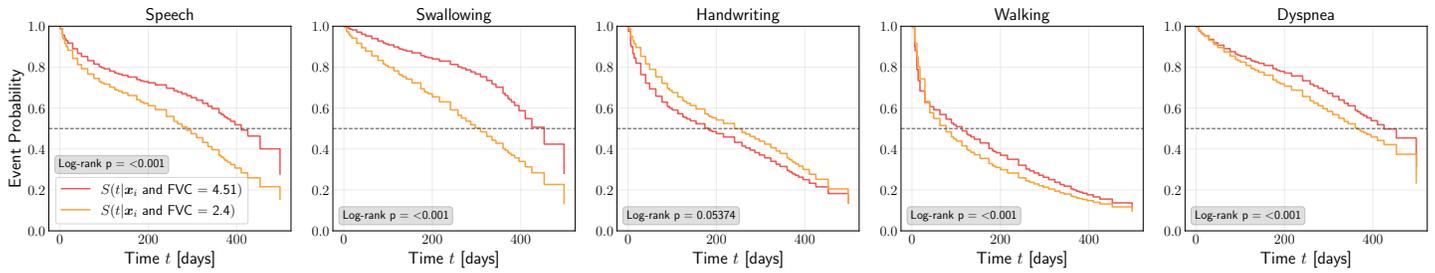

\begin{adjustwidth}{-2.25in}{0in}
\centering
\textbf{[Fig. 7]}
\caption{Predicted ISDs for Mr. Smith based on his FVC score being high or low.}
\label{fig:proact_isd_fvc}
\end{adjustwidth}
\end{figure}

Next, forced Vital Capacity (FVC) is a measure of lung function and represents the total amount of air a person can forcibly exhale after taking a deep breath. It is commonly used to assess lung respiratory function and is measured in liters. In ALS, FVC is a key clinical assessment for evaluating respiratory muscle strength, which progressively weakens as the disease progresses. Mr. Smith's total FVC score at baseline was 2.69. Again, we ask the counterfactual question of what if instead he had a higher FVC or a lower FVC. Fig. \ref{fig:proact_isd_fvc} shows the predicted ISDs for Mr. Smith based on whether he has a high or low FVC. As expected, the \emph{Handwriting} event shows only a weak relationship with FVC, as evidenced by the high $p$-value and the alignment of the two survival curves. In contrast, the \emph{Swallowing} event exhibits a significant relationship with FVC, and the model provides substantially different time-to-event predictions based solely on this covariate.

\begin{figure}[!ht]
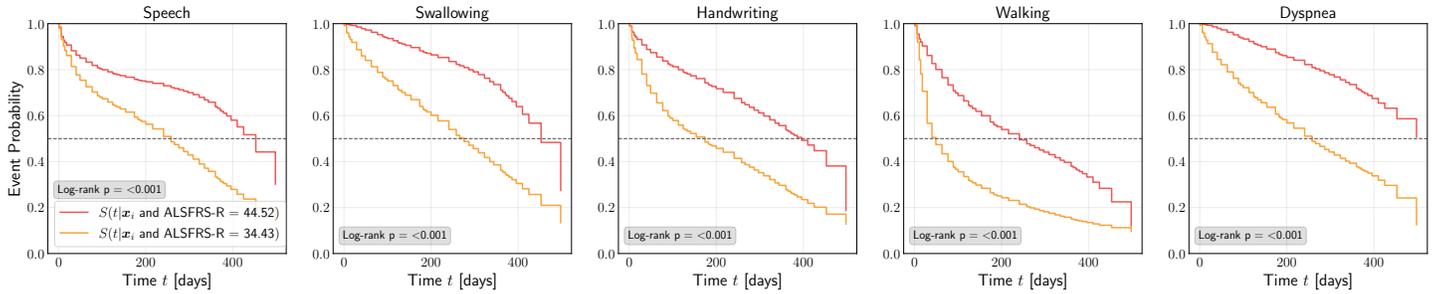

\begin{adjustwidth}{-2.25in}{0in}
\centering
\textbf{[Fig. 8]}
\caption{Predicted ISDs for Mr. Smith based on whether his initial ALSFRS-R score is high or low.}
\label{fig:proact_isd_alsfrs}
\end{adjustwidth}
\end{figure}

Lastly, we consider the total ALSFRS-R score at baseline and its effect on predicted functional decline. This score provides a snapshot of Mr. Smith's initial physical capabilities when he is relatively healthy before the disease has advanced notably. Mr. Smith's total ALSFRS-R score at baseline was 37. Again, we make a counterfactual prediction if instead Mr. Smith had a significantly higher or lower ALSFRS-R score. Fig. \ref{fig:proact_isd_alsfrs} shows the predicted ISDs for Mr. Smith based on whether he has a high or low ALSFRS-R score. Consistent with the literature~\cite{Kjaeldgaard2021}, if Mr. Smith had a high ALSFRS-R score to begin with, he would (presumably) experience functional decline much later than if he had a low ALSFRS-R score. This is indicated by the discrepancy between the survival curves and the low $p$-values across all the events. Obviously, initial physical ability has a strong relationship with loss of function after a certain time, and patients may be at different stages of the disease and have different levels of strength when they enter the trial. In addition, research in ALS has suggested the presence of various disease phenotypes with different progression rates~\cite{Swinnen2014}.

\section*{Discussion}

\subsection*{Contribution}

This research has three goals: (1) determine whether covariate-based survival models can predict the time to individual functional decline (ALSFRS-R $\leq$ 2) in ALS patients more accurately than population estimators, (2) precisely estimate risk scores and event times, and (3) explore how counterfactual scenarios affect the probability of functional decline over the course of the disease.

For the first goal, individualized predictions are crucial for tailoring healthcare and therapeutic decisions to the individual patient. Models that estimate an individual's risk rather than a group average enable us to answer important questions from the patient's point of view such as: how much longer will I be able to speak clearly enough to be understood? We make a novel contribution in this regard, as we empirically show that most covariate-based models, even simple linear models, outperform the KM estimator (which does not use covariates) in predicting the time to functional decline. Given that ALS affects multiple functional systems (bulbar, motor, respiratory) at varying rates, our analysis also shows clear advantages of adopting non-proportional hazards models, such as Random Survival Forests and MENSA. Unlike other methods for predicting functional decline in ALS~\cite{Ong2017, Vieira2022, Amaral2024, Jabbar2024, Turabieh2024}, our method uniquely provides patients and clinicians with an actual time (in days) when functional decline is likely to occur.

For the second goal, we rigorously evaluate five survival models across three aspects of model performance: discrimination, absolute/squared error, and calibration. The main takeaway is that some types of functional decline in ALS patients are easier to predict than others. For example, the \emph{Speech} event has an average Harrell's CI of approximately 74\% across the five models -- the highest among all events -- but also an average mMAE (a variant of the MAE that supports censoring) of 192.4 days, the second-highest. This means that even if a model is good at ranking patients by their relative risk (independent of time) for some event, it cannot necessarily predict accurately when the event will occur. In contrast, the \emph{Walking} event has a lower average CI of approximately 67\%, but an average mMAE of only 147.7 days. This suggests that, although it is harder to rank patients by risk for this event compared to \emph{Speech}, it is easier, by contrast, to predict when the event will occur. This leads to the realization that strong ranking performance does not necessarily translate into accurate time-to-event predictions.

Finally, for the third goal, we use our method to make counterfactual predictions, investigating how modifying specific covariates affects predicted disease outcomes for individual patients. To our knowledge, no prior work has explored this use case in ALS. However, such predictions are highly relevant for medical professionals who must make important treatment decisions after first meeting a patient. For example, consider our patient, Mr. Smith: based on his covariates, there were no significant differences in the predicted survival probabilities (\ie the risk of functional decline) should he decide to take Riluzole, a drug used to treat ALS. In other words, whether he decided to take Riluzole or not, this decision would not significantly change his predicted risk of functional decline over the next 500 days from the first time he saw the neurologist. We believe that these individualized predictions can be very beneficial for treatment planning and clinical decision-making.

\subsection*{Application}

Our method has several important applications. First, it can help medical doctors make informed decisions by tailoring treatments and interventions to the individual's predicted disease trajectory. This is crucial in ALS, where timely interventions, such as assistive devices and nutritional or respiratory support, can significantly impact a patient's quality of life. Second, patients and their families are often faced with difficult decisions regarding treatment options, lifestyle adjustments, and end-of-life planning. By providing a clear, individualized prognosis, clinicians can help patients make more informed decisions about their care, improve their quality of life, and give them a sense of control over their treatment. As part of the evaluation, we show how our method can make counterfactual predictions, answering the what-if questions: Mr. Smith was diagnosed with limb-onset ALS and the model predicts a specific path of functional decline for him, but what would have happened if he had been diagnosed with bulbar-onset ALS instead. Third, personalized predictions can also help design and evaluate clinical trials. Understanding the likely timing of functional decline in patients with ALS can help researchers identify suitable candidates for clinical studies, ensuring that the trials are conducted with participants at the appropriate stage of disease. This can lead to more effective testing of potential therapies and interventions that could slow or stop disease progression.

Concretely, if practitioners wish to implement our method in everyday clinical practice, we recommend the following steps:

\begin{enumerate}
    \item Integrate model training with existing electronic health record (EHR) systems or clinical decision support tools to automatically retrieve patient data (covariates) and past ALSFRS-R scores (event information). After obtaining the relevant data, they should train the model, evaluate it, deploy it, and expose an application programming interface (API) that allows clinicians to make predictions on new patients.
    \item Develop a user-friendly software interface or web-based application that allows clinicians to easily input or access patient information and obtain individual predictions of functional decline over time. When a new patient is diagnosed with ALS and has their first visit at the hospital, clinicians can use the application to enter the patient's covariates and obtain a prediction over the next 500 days.
    \item Update, retrain, and deploy the model regularly with new patient data to maintain accuracy and incorporate new biomarkers or additional clinical variables, if such become available. This can be done using Machine Learning Operations (MLOps), which is a set of practices that automate and simplify machine learning workflows and deployments.
\end{enumerate}

\subsection*{Limitations}

All survival models evaluated in this work assume conditional independent censoring, \ie that the event time is independent of the censoring time, given the patient's covariates. For example, patients who leave the study early do so for reasons unrelated to the event of interest or because of poor lung function captured by the forced vital capacity covariate. However, ALS is a heterogeneous disease that can quickly deteriorate and no study can capture all intrinsic disease mechanisms. Generally, clinical trials in ALS experience high dropout rates~\cite{Atassi2013}, which is also evident in our study through the high censoring rates shown in Table~\ref{tab:dataset_overview} and Fig.~\ref{fig:event_distribution}. It is likely that this phenomenon is related to disease progression and therefore functional decline, but since we do not observe functional decline in patients who leave the study, this kind of censoring may violate the assumption of conditional independent censoring and bias the results. According to Atassi et al.~\cite{Atassi2013}, the three most common reasons for attrition (\ie dropout) in ALS trials were withdrawal of consent for no specific reason (37\%), death (28\%), and withdrawal secondary to an adverse event (17\%).

\subsection*{Future work}

Our method has several avenues for future research. First, addressing the aforementioned dependency between the event and censoring times is relevant, given the relatively few covariates provided in the PRO-ACT dataset and because all evaluated survival models assume that censoring is conditionally independent. To address this issue, when estimating model parameters, we can learn a so-called copula function~\cite{Emura2018}. This is a link function that describes the dependency structure between the event and censoring distributions. Copulas have recently gained attention in survival analysis due to their flexibility in capturing complex dependency structures~\cite{foomani_copula-based_2023}. Second, it is highly relevant to include covariates captured using magnetic resonance imaging (MRI) into the predictive models, since such features can explain early signs of neurodegeneration, for example, cerebral atrophy in the motor cortex and corticospinal tract~\cite{Dadar2020}. At present, such features have only been used to predict mortality or disease aggressiveness in patients with ALS~\cite{Dadar2020, Lajoie2024, Dieckmann2022}, but not to predict functional decline. Thus, this presents an exciting avenue for future research.

\section*{Conclusion}

This article proposes a novel and interpretable method for estimating the time to functional decline in ALS patients. We formulate this task as a multi-event survival problem and validate our method in the PRO-ACT dataset by training five covariate-based survival models to predict significant functional impairment with respect to five common functions. The proposed method has the following advantages over state-of-the-art approaches: (1) It explicitly predicts the time until functional decline is likely to occur for an individual patient. (2) It gives superior time-to-event estimates over commonly used population estimators, such as the KM estimator. (3) It allows for individual counterfactual predictions, where certain covariates can be changed to see their effect on the predicted outcome. In conclusion, we recommend using the proposed method on patient records in clinical settings, subject to additional rigorous validation, to assess the risk of imminent functional decline and enable more personalized treatment planning.

\section*{Supporting information}

\paragraph*{S1 Table.}
\label{S1_Table}
\textbf{List of covariates.}\\
\noindent (DOCX)

\paragraph*{S2 Table.}
\label{S2_Table}
\textbf{List of missing data rows.}\\
\noindent (DOCX)

\paragraph*{S3 Table.}
\label{S3_Table}
\textbf{List of model hyperparameters.}\\
\noindent (DOCX)

\paragraph*{S4 Table.}
\label{S4_Table}
\textbf{Mr. Smith's covariates.}\\
\noindent (DOCX)

\section*{Acknowledgments}

Data used in the preparation of this article were obtained from the Pooled Resource Open-Access ALS Clinical Trials (PRO-ACT) Database. The following organizations and individuals within the PRO-ACT Consortium contributed to the design and implementation of the PRO-ACT Database and/or provided data, but did not participate in the analysis of the data or the writing of this report: ALS Therapy Alliance, Cytokinetics, Inc., Amylyx Pharmaceuticals, Inc., Knopp Biosciences, Neuraltus Pharmaceuticals, Inc., Neurological Clinical Research Institute (MGH), Northeast ALS Consortium, Novartis, Orion Corporation, Prize4Life Israel, Regeneron Pharmaceuticals, Inc., Sanofi, Teva Pharmaceutical Industries, Ltd., and The ALS Association.

\nolinenumbers

\end{document}